\documentstyle[aps,prl,multicol]{revtex}
\begin{document}
\draft
\title{
%To be or not to be localized:\\ 
Localization of fermions in an anisotropic random magnetic field}

\author{Yong Baek Kim}
\address{
Department of Physics, The Pennsylvania State University,
University Park, PA 16802, U.S.A.\\
Institute for Theoretical Physics, University of 
California, Santa Barbara, CA 93106, U.S.A.\\
}

%\date{Received: November 28, 1998; Revised: January 27, 1999}
\maketitle

\begin{abstract}

We study the localization of fermions in an anisotropic random 
magnetic field in two dimensions.
It is assumed that the randomness in a particular direction
is stronger than those in the other directions.
We consider a network model of zero field contours, 
where there are two types of randomness - the random 
tunneling matrix element at the saddle points and unidirectional 
random variation of the number of fermionic states following zero 
field contours.
After averaging over the random complex tunneling amplitude, 
the problem is mapped
to an $SU(2N)$ random exchange quantum spin chain in the
$N \rightarrow 0$ limit.
We suggest that the fermionic state becomes critical
in an anisotropic fashion.

\end{abstract}


\begin{multicols}{2}

The localization problem of a nonrelativistic particle in
a random perpendicular magnetic field has been a subject of
intensive research and debate.
In particular, the problem arises from the study of
disordered half-filled Landau level in the quantum
Hall regime\cite{HLR}. 
In the absence of disorder, the Chern-Simons
theory of electrons in the half-filled Landau level leads to 
an effective theory of composite fermions in the fluctuating 
gauge field\cite{HLR}.
Due to the fact that the density distribution of the
fermions are tied to the fluctuating magnetic field,
static disorder leads naturally to the random static
magnetic field with zero mean\cite{KZ}. 
Since the longitudinal resistivity of the
electrons in the half-filled Landau level is the same
as that of composite fermions, the problem of localization
of composite fermions in a random static magnetic field
is important for understanding transport properties at 
the half-filling.

One might expect that the system belongs to the unitary universality
class as the disordered system with broken time reversal
symmetry and zero Hall conductance. According to conventional 
scaling theory, all the states are localized in two dimensions
for the unitary class\cite{Lee}.
It is established that those systems can be described by 
a unitary non-linear sigma model (NL$\sigma$M) {\it without 
topological term} in the long wavelength limit.
Some years ago, Zhang and Arovas\cite{ZA} 
suggested that local fluctuation
of $\sigma_{xy}$ can induce the long-range interactions between
topological densities in the NL$\sigma$M. It was claimed that
this may give rise to a novel delocalization transition
when the conductance at short distances becomes larger than
a critical value of the order of $e^2/h$.

Since then, it has been controversial whether the model
has a delocalized state or not. 
There have been several numerical and analytical studies
leading to different conclusions.
Some of them conclude that the system indeed belongs to
the unitary class and all the states are 
localized\cite{derek,kim,nagaosa}.
On the other hand, some numerical calculations on the
lattice model reached the conclusion that there are 
delocalized states near the band 
center\cite{ZA,liu}.
The very recent extensive 
numerical calculation by Furusaki\cite{furusaki} 
showed that a state in the band center is not localized
when there is a special particle-hole symmetry
in the lattice model.
On the other hand, in the absence of the particle-hole
symmetry, all the states are localized. This suggests
that there is no extended state in the generic
continuum model\cite{furusaki}. 

Previously Furusaki, Lee, and the author\cite{kim} 
map a network model\cite{derek} of snake states 
(which are fermionic states 
following zero field contours) to a quantum spin chain.
They suggest that the quantum spin chain has a gap
in the excitation spectrum so that the fermions in
the original problem is always localized.
In this paper, we elaborate this mapping further and
point out that some ideas of Zhang-Arovas theory may
be realized in an anisotropic version of the random
flux problem. In particular, we suggest that the
number of snake states can fluctuate in space, which 
was set to be a fixed number in previous numerical
and analytical studies. We show that if the number of
snake states changes only in a particular
direction, the fermionic state becomes {\it critical}
and the critical behavior could be different
for different directions.

Let us first review the mapping from a network model
of snake states to a quantum spin chain\cite{kim,dhlee}.
We assume that the random magnetic field has zero
mean and a non-zero variance.
We consider the semiclassical limit in which the 
correlation length of the random magnetic field is
large compared to the typical magnetic length.
In this limit of smooth disorder, a fermion moves
along contours of constant magnetic field. 
The contours which percolate geometrically across
the system would be $B=0$ contours. It turns out that
semiclassical motion of a fermion along $B=0$
contours follows snake-like trajectories.
In the network model, these zero field contours are
represented as links on a square network, carrying 
the fermionic ``snake'' states\cite{derek}.
The node of the network corresponds to the saddle
points of the magnetic field where the scattering
to either the right or left occurs\cite{derek}.
The quantum interference due to these scattering
processes are modeled by random tunneling 
amplitudes at the saddle points and the mixing
of the snake states.

It is important to notice that the fermionic states
along the zero field contours arise in pairs at a 
given energy and they are propagating in the same
direction. This is because these states are generated
from a symmetric and antisymmetric linear combinations
of the Landau levels on either side of the contour.
This implies that there exist two snake states per 
each local Landau level below the Fermi energy.
Therefore, there are always even number, $M$, of snake
states in each link of the network model.
Previous studies assumed that the number of snake
states, $M$, is fixed\cite{derek,kim}. 
However, notice that $M$ could
be a spatially varying quantity as far as this is an even
number because the magnitude of the magnetic field
changes in space. Later, we argue that if the change
of $M$ occurs only in one direction, then it leads
to a significant effect. At the moment, we assume that 
$M$ is a fixed even number.
 
In the network model, two types of random complex 
tunneling amplitude are introduced, $t_j$ 
and $t_{jk}$. $t_j$ represents the
tunneling between the snake states, $\psi_j$ ($j=1,...,M$), 
in neighboring zero field contours and $t_{jk}$ 
corresponds to the mixing between snake states in different 
channels (among $M$ number of channels) in the same 
zero field contour.
Using the replica trick, the disorder averages over $t_j$ 
and $t_{jk}$ are done. Taking $y$ as the imaginary time,
one can regard the disorder averaged action as representing 
interacting fermions, $\psi_{\alpha j}$ ($j=1,...,M$ and 
$\alpha=1,...,2N$) on $M$ number of coupled one dimensional 
systems. Here $N$ is the replica index.
Substituting the fermion operators with the generators of
$SU(2N)$:
\begin{equation}
S^{\alpha}_{j,\beta} = \psi^{\dagger}_{\alpha j} 
\psi_{\beta j} - \delta_{\alpha \beta} {1 \over 2N}
\sum_{\gamma} \psi^{\dagger}_{\gamma j} \psi_{\gamma j} \ ,
\end{equation} 
we can obtain $M$ coupled $SU(2N)$ spin chains given by
the Hamiltonian\cite{kim}:
\begin{eqnarray}
H &=& \sum_{x} \sum_{j} J_j {\rm Tr} [S_j (x+a) S_j (x)] \cr
&&+ \sum_{x} \sum_{j < k} J_{jk} {\rm Tr} [S_j (x) S_k (x)] \ , 
\end{eqnarray} 
where $a$ is the lattice spacing corresponding to typical
distance between snake snakes.
Here $J_j = \langle |t_j|^2 \rangle$, $J_{jk} = 
-\langle |t_{jk}|^2 \rangle$, and 
${\rm Tr}[A B] = \sum_{\alpha \beta} A^{\beta}_{\alpha}
B^{\alpha}_{\beta}$ for any $2N \times 2N$ matrix $A$ and $B$.
In the case of decoupled chains, each spin chain is described  
by a particular representation of $SU(2N)$, characterized
by a Young tableau with a single column of length $N$.
 
Notice that there are antiferromagnetic couplings ($J_j > 0$)
in each spin chain $j$ and ferromagnetic couplings ($J_{jk} < 0$)
between the spin chain $j$ and $k$ for each $x$.
The spin at each site represents a snake state at 
a given time at that particular position.
The antiferromagnetic coupling, $J_j$, comes from the
fact that the tunneling between the snake states in
the same channel always occurs between oppositely propagating
snake states. On the other hand, the mixing between snake
states in different channels happens for snake states
propagating in the same direction, which leads to the 
ferromagnetic coupling, $J_{jk}$.
Notice that each chain corresponds to a single-channel
network model with non-zero Hall conductance.
In order to impose the constraint of zero Hall conductance
we require the alternation of the bond strength on
the neighboring chains to be staggered.
This means that $J_j (x={\rm even}) = J_{j+1} (x={\rm odd})$
and $J_j (x={\rm odd}) = J_{j+1} (x={\rm even})$.
In terms of the original problem, the localization of
the fermions corresponds to the existence of an energy gap
in the spin chain (in the replica limit, $N \rightarrow 0$).

Due to the fact that there are $M$ number of snake 
states in each link, the conductance of the system at
short distances is $G = M {e^2 \over h}$.
Thus the large conductance limit, $M \gg 2$, corresponds
to the large number of coupled chains.
The interchain coupling is ferromagnetic and it exists 
for any two of $M$ spin chains. Thus, as $M$ becomes large
we expect that, the
coupled chains become essentially single spin chain
with a uniform antiferromagnetic coupling ${\cal J}$\cite{kim}:
\begin{equation}
H = \sum_{x} {\cal J} {\rm Tr} [S (x+a) S (x)] \ ,
\end{equation}
where $S (x)$ is in the totally 
symmetric representation described by a 
Young tableau
with $M$ number of columns of length $N$.
It is useful to think about the $SU(2)$ case ($N=1$).
In this case, $M$ corresponds to $2S$, where $S$ is
the ordinary spin quantum number. Since $M$ is
always an even number, the system can be described
by an integer spin chain. Therefore, there is a gap
in the excitation spectrum in the spin chain. If one 
assumes that $N \rightarrow 0$ limit is not singular, 
this may imply that the electrons are localized in 
the original problem even for large bare conductances.

Now let us consider the case that the number of snake
states is not fixed, but can change in space.
In particular, we assume that the number of snake
states changes only in one direction.
We will discuss the isotropic case later.
In this anisotropic random flux problem, we can take
the $x$ direction as the direction in which $M$
changes in space. We assume that there are two types
of randomness in the problem - one of them is the
random tunneling amplitude at the saddle points and
the other one is the number of snake states.
After performing the average over the random tunneling
amplitude, we get 
\begin{eqnarray}
H &=& \sum_{p} \sum_{l_p \le x \le r_p} 
\sum_{1 \le j \le M_p} J_j {\rm Tr} [S_j (x+a) S_j (x)] \cr 
&&+ \sum_{p} \sum_{1 \le j \le {\rm Min} (M_p, M_{p+1})} 
J_j {\rm Tr} [S_j (l_{p+1}) S_j (r_p)] \cr 
&&+ \sum_{p} \sum_{l_p \le x \le r_p} 
\sum_{1 \le j < k \le M_p} 
J_{jk} {\rm Tr} [S_j (x) S_k (x)] \ .
\end{eqnarray}
Here the coupled spin chains consist of many subsystems
labeled by $p$.
$l_p$ and $r_p$ are the positions of the left and 
right ends of the $p$th subsystem ($l_{p+1} = r_p + a$). 
Each subsystem labeled by $p$ has $M_p$ number of
coupled spin chains. In general, $M_{p+1} - M_p = M^{'}_p$ 
is an even number.   
In the large bare conductance limit, where $M_p$ is much 
larger than 2, the effective Hamiltonian can be
written as
\begin{eqnarray}
H &=& \sum_{p} \sum_{l_p \le x \le r_p} {\cal J}_p 
{\rm Tr} [S_p (x+a) S_p (x)] \cr 
&&+ \sum_{p} {\rm Min} ({\cal J}_p, {\cal J}_{p+1}) 
{\rm Tr} [S_{p+1} (l_{p+1}) S_p (r_p)] \ .
\label{effchain}  
\end{eqnarray}
Here the system still consists of many subsystems.
Each subsystem labeled by $p$ is a uniform antiferromagnetic 
spin chain in a representation described by a Young tableau
with $M_p$ number of columns of length $N$.

It is instructive to think about $SU(2)$ case ($N=2$).
In this case, the above Hamiltonian represents the coupled
antiferromagnetic integer spin chains. Each integer spin
chain has a uniform antiferromagnetic coupling ${\cal J}_p$ 
and the spin magnitude $S_p = M_p/2$. Thus each subsystem
would have a gap in the excitation spectrum if it were an 
infinite system. Notice that the second term in the Hamiltonian
corresponds to the coupling between edge spins of adjacent
spin chains. 
For an integer spin chain with finite size and
spin magnitude $S_p$, the edge spin has the spin magnitude
$S_p/2$\cite{ng}. 
This can be seen from the Berry phase contribution
to the effective action for an integer spin chain of length $L$:
\begin{eqnarray}
S_{\rm BP} &=& S_p \sum_{x} (-1)^{x/a} \Omega (x) \sim 
{S_p \over 2} \int dx {\partial \Omega (x) \over \partial x} \cr
&&= {S_p \over 2} \left [ [\Omega (L) - \Omega (0) ] 
+ 4 \pi Q \right ] \ ,
\end{eqnarray}
where $\Omega (x)$ is the solid angle subtended by the
closed path on the surface of a unit sphere defined
by the time evolution of a unit vector at $x$.
$Q$ is an integer measuring the number of times the 
spin configuration covers the surface of the sphere.
Since ${S_p \over 2} \times 4 \pi Q = 2 \pi Q$, the last
term has no effect and can be dropped.

At the interface $x = L$, between the spin chains with 
$S_p$ and $S_{p+1}$, the Berry phase contribution in the
limit $L \gg a$ becomes
\begin{equation}
S_{\rm BP} = {S_p \over 2} \Omega (L) - 
{S_{p+1} \over 2} \Omega (L+a) \approx 
{S_p - S_{p+1} \over 2} \Omega (L) \ .
\end{equation}
Since $(S_p - S_{p+1})/2 = (M_p - M_{p+1})/4$, the edge
spin can be both the half-integer and integer spin.
In the system described by Eq.(\ref{effchain}), this generates
edge spins with random spin magnitude $S^{\rm edge}_p (X)$ 
at the interface, $X$, between adjacent $p$th and $(p+1)$th
integer spin chains. Furthermore, depending on 
whether $r_p-l_p$ is even or odd, the residual
couplings between those edge spins can be either
antiferromagnetic or ferromagnetic. The magnitude
of the coupling goes as ${\cal J}_p e^{-(r_p-l_p)/\xi_p}$,
where $\xi_p$ is the correlation length.
From the above consideration, one can see that the
effective coarse-grained Hamiltonian of the system 
in the case of $SU(2)$ is given by
\begin{equation}
H = \sum_{X} {\cal J}(X) {\bf S}^{\rm edge}(X+A) \cdot 
{\bf S}^{\rm edge}(X) \ ,
\label{su2}
\end{equation}
where ${\cal J}$ has a distribution of both antiferromagnetic
(${\cal J} > 0$) and ferromagnetic couplings (${\cal J} < 0$).
Here the spin magnitude can be also random in space and $A$
is the typical distance between interfaces of different
subchains.

Coming back to $SU(2N)$ case, one can generalize the
above argument. For example, at the interface, $X$, between 
two adjacent spin chains, the edge spin is generated and it is 
in the representation characterized by a Young tableau with
$M^{\rm edge} = ({\rm even \ number})/2$ number of columns of length $N$.
At the same time, the sign and magnitude of the residual 
spin exchange interaction between those edge spins can be
random. Thus the effective Hamiltonian can be written as
\begin{equation}
H = \sum_{X} {\cal J}(X) {\rm Tr} [S^{\rm edge}(X+A) 
S^{\rm edge}(X)] \ .
\end{equation}
The number of columns of the Young tableau for the
edge spins at $X$ has a random distribution.

It is rather easy to understand what the effective Hamiltonian
stands for in the original snake-states picture.
In the region where the number of snake states changes from
one even number to the other, the number of remaining snake
states after forming all possible localized configuration
out of available snake states, is still always even and
they are all in the same direction. 
Since most of the snake states in the bulk are localized,
the remaining degrees of freedom are described by
the weak coupling across a long distance 
between residual snake states at the interface between
two region of different number of snake states.
The number of residual snake states can be an arbitrary
even number. 
%On the other hand, the residual weak coupling
%across a long distance can occur between residual snake states 
%propagating in either the same or opposite directions.
In the limit of strong mixing, the residual snake states
can be represented by a single large snake state.
In the spin chain language, the residual snake
states correspond to $S^{\rm edge}$. Thus the number of
columns in the Young tableau corresponds to the number
of residual snake states at the interface. 
If the weak coupling occurs between the residual
snake states propagating in the same direction, the coupling
in the spin chain is ferromagnetic. For the snake states
propagating in the opposite direction, the coupling 
becomes antiferromagnetic.

The $SU(2)$ case described by the Hamiltonian Eq.(\ref{su2})
was studied by various methods\cite{eric,sigrist}. 
It turns out that the real space
renormalization group (RG)\cite{dasgupta,fisher} 
treatment can provide a lot of insight 
about the ground state of the system\cite{eric}.
Let us consider an isolated link with two sites, where two 
neighboring spins are coupled by ${\cal J}$.
The ground state of an isolated two-site system would have
maximum (${\cal J} < 0$) or minimum (${\cal J} > 0$) spin 
quantum number depending on the sign of the exchange coupling.
There would be an energy gap $\Delta$ to the first excited
state. Since the total system consists of these links, the
system can be described by a distribution of energy gaps 
and the spin magnitude. In the real space RG scheme, 
one successively replaces the strongest links, where 
the energy gap is the largest, by effective spins. 
The RG transformation preserves the form of the Hamiltonian,
but it changes the probability distribution of the
gaps (or exchange couplings) and the spin magnitude.
%Thus RG flows exist in the space of distributions of gaps
%and spin magnitude. 

In the low energy limit, the RG flows
reach a single universal fixed point and the distributions
become universal at the fixed point\cite{eric}.
At low energies, the random spin chain can be described 
by weakly coupled large effective spins\cite{eric}. 
%Here the large spin corresponds to a correlated segment.
As the links are replaced by effective spins, the effective
couplings or the gaps of the links decrease.
At the same time, the average distance between neighboring
spins and the magnitude of the effective spins increase.
It was found that the average size of the effective spins
diverges as $T^{-\alpha}$ while the average distance
between neighboring effective spins increases 
as $T^{-2\alpha}$\cite{eric}. 
That is, the correlation length diverges
in the zero temperature limit.
This suggests that the following correlation function decays
as a power law at the zero temperature\cite{eric}.
\begin{equation}
C (X-Y) = \langle {\cal E}_{XY} {\bf S} (X) \cdot {\bf S} (Y)
\rangle \propto {1 \over |X-Y|^{\nu}} \ ,
\end{equation}
where ${\cal E}_{XY} = \Pi_{Z=X}^{Y-1} 
{\rm sgn} (-{\cal J}(Z))$.

In the original problem of fermions in an anisotropic 
random flux distribution, this may imply that the fermionic
state is {\it critical} in the direction of changing number
of snake states or large variation of the magnitude of 
random magnetic field. 
This means that the wavefunction is not exponentially
localized in this direction, but decays as a power law
as a function of system size.
Therefore, in some sense, it behaves like a critical
state at the transition between metal and insulator. 
The reason why the correlation becomes critical is that 
some rare events across the system in the $x$-direction
dominates the disorder averaged correlation function.
Since the spin chain does not have an excitation gap, the 
disorder averaged correlation in the time direction cannot 
decay exponentially. Due to the critical nature of the
spin chain, we expect that it decays also 
as a power law even though the power law would be 
in general different for different directions. 
Thus, the fermionic state in the 
$y$-direction is also critical and the
critical behavior may be different from that in the
$x$-direction. In the absence of detailed information
about the exponents $\nu$ for different directions, 
it is hard to tell in which direction the wavefunction
decays faster.
%The typical correlation still shows 
%exponentially-decaying behavior.
%The rare events or correlations occur between the spins
%separated in space (in the $x$-direction). 
%Therefore, the correlation in the $x$-direction becomes
%power-law while the correlation in the imaginary time
%(or the $y$-direction) remains to have 
%exponentially-decaying behavior.
%Thus we expect that the electronic states are exponentially
%localized in the $y$-direction as the usual unitary
%case. 

%One can see that the motion of fermions in the 
%direction of changing number of snake states are 
%highly correlated. This is very similar
%in spirit to Zhang-Arovas theory in the sense that 
%there exists a long range interaction between 
%topological densities which may corresponds to 
%highly correlated large snake states.
%However, this happens in the anisotropic flux 
%problem. 
In the isotropic problem, the number of
snake states can change in any direction.
In the spin chain language, it will induce the
time and space dependent random spin size and 
the couplings. Thus one gets an annealed disorder
problem instead of a quenched one. 
Even though it is not clear what the effects of this type of
disorder would be, it is possible that the
effects of annealed disorder could be much weaker
so that the random-tunneling-only model may be enough
to describe the isotropic random flux problem.   

In conclusion, we consider the motion of fermions
in an anisotropic random magnetic field.
We assume that the number of fermionic states following
zero field contours can be changed only in one
direction. The mapping from a corresponding network
model to a random quantum spin leads to the conclusion
that the fermionic states are critical in an anisotropic
fashion.

We thank Akira Furusaki, Ilya Gruzberg, Steve Kivelson, 
T. Senthil and especially Shou-Cheng Zhang for 
helpful discussions.
The research at ITP, UCSB was supported in part by
NSF grant No. PHY9407194. This work was also 
supported by Alfred P. Sloan Foundation Fellowship.

\end{multicols}

\end{document}